\documentclass[aps,prl,twocolumn,showpacs,superscriptaddress,groupedaddress,amsmath,amssymb]{revtex4-1}  
\usepackage{graphicx}  
\usepackage{subfigure}
\usepackage{braket}
\usepackage{longtable}
\usepackage{bbold}
\usepackage{hyperref}
\usepackage{url}
\hypersetup{colorlinks,breaklinks,
citecolor=[rgb]{0.0,0.5,0.5},
    urlcolor=[rgb]{0.0,0.5,0.5},
    linkcolor=[rgb]{0.0,0.5,0.5}}
\DeclareMathOperator{\Tr}{Tr}

\begin{document}

\title{Qutrit Magic State Distillation Tight in Some Directions}
\author{Hillary Dawkins$^1$ and Mark Howard$^2$}
\affiliation{%
Institute for Quantum Computing$^{1,2}$, Department of Physics and Astronomy$^1$ and Department of Applied Mathematics$^2$,
University of Waterloo, Waterloo, Ontario, Canada, N2L 3G1
}


\begin{abstract}

Magic state distillation is a crucial component in the leading approaches to implementing universal fault-tolerant quantum computation, with existing protocols for both qubit and higher dimensional systems. Early work focused on determining the region of distillable states for qubit protocols, yet comparatively little is known about which states can be distilled and with what distillable region for $d>2$. Here we focus on $d=3$ and present new four-qutrit distillation schemes that improve upon the known distillable region, and achieve distillation tight to the boundary of undistillable states for some classes of state. As a consequence of recent results, this implies that there is a family of quantum states that enable universality if and only if they exhibit contextuality with respect to stabilizer measurements. We also identify a new routine whose fixed point is a magic state with maximal sum-negativity i.e., it is maximally non-stabilizer in a specific sense. 
\end{abstract}

\maketitle

One of the main challenges facing the implementation of a large scale quantum computer is the ability to protect quantum information from decoherence, typically introduced by unwanted interactions with the environment. Encoding of information in quantum error correcting codes provides a partial answer \cite{Shor,Kit,Raus}. However such encodings normally only allow a limited set of transversal or manifestly fault-tolerant operations, usually the stabilizer operations i.e., Clifford gates, preparation of stabilizer states and Pauli measurements. Stabilizer operations alone do not enable universality \cite{Gottesman1} and therefore some additional resource will be required to supplement them in any proposal for universal fault-tolerant quantum computation.

\begin{figure}
\includegraphics[scale=0.85]{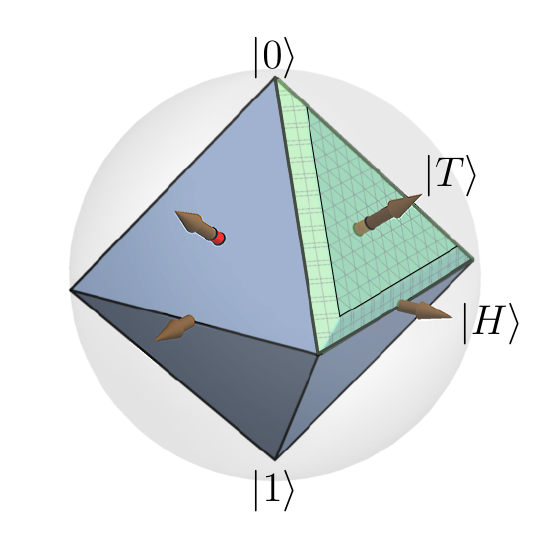}
\caption{\label{fig:BK} The qubit Bloch sphere and the stabilizer octahedron: The convex hull of stabilizer eigenstates -- the stabilizer polytope -- carves out a solid octahedron. $H$-type magic states lie outside the octahedron edges, and $T$-type magic states lie outside the octahedron faces. The green volume outside one face represents the region in which ancillas have not yet been shown to be distillable.}
\end{figure}

A leading solution (although alternatives exist \cite{Paetznick:2013,TJOC:2013}) is provided by the magic state distillation protocol, first proposed by Knill \cite{Knill:2005} and Bravyi and Kitaev \cite{BK}. There it was shown that stabilizer operations may be promoted to universal fault-tolerant quantum computation when supplemented by a supply of an additional resource state, known as a magic state. 
Furthermore, these magic states may be prepared through an iterative procedure in which less pure states are consumed to produce a higher fidelity output state using only stabilizer operations. If the input ancillas to a distillation routine are expressible as a mixture of stabilizer states (geometrically, if they are inside the stabilizer polytope e.g.~Fig.~\ref{fig:BK}) then no amount of stabilizer operations can produce a magic state. 
 Reichardt \cite{Reichardt} showed that the distillable region was tight to the stabilizer boundary along the octahedron edges in the so-called $H$-direction (see Fig.~\ref{fig:BK}), while Campbell and Browne \cite{CB1,CB2} showed that (for stabilizer codes of fixed length) there exists a region of undistillable non-stabilizer states outside the octahedron faces in the $T$-direction. 

For qudits (hereafter this refers to systems of odd prime dimension $d$), 
undistillability of a state is implied by it having a quasi-probabilty representation (in Gross' discrete Wigner function \cite{Wootters:1987,Gross:2006}) that is everywhere non-negative \cite{Victor}. Conversely, there is currently no known impediment to the distillability of states that do exhibit negativity -- it has been conjectured to be a sufficient condition \cite{Victor2}. Recently, the presence of negativity was shown to exactly coincide with the possibility of exhibiting state-dependent quantum contextuality using stabilizer measurements \cite{Context}.
Proving that the presence of negativity (or contextuality) is sufficient to boost stabilizer operations to universality requires showing that any state that is negatively represented is distillable by some magic state distillation routine. While magic state distillation protocols for all prime dimensions \cite{CAB1,CAB2,CAB3} (some highly efficient \cite{CAB2}) have been found, no single protocol has been shown to distill states tight to the undistillable boundary (note that tight distillation was previously achieved for a single direction using a combination of stabilizer codes in \cite{CAB1} however). Here we focus on $d =3$ and present distillation schemes that achieve tight distillation for multiple directions using less demanding protocols, as well as showing distillability of a geometrically significant and maximally non-stabilizer magic state.

\textit{Definitions.$-$} We begin with a brief overview of the magic state distillation protocol and provide some useful definitions. Magic state distillation procedures are based on stabilizer error correcting codes, which can be described by a set of generators $\{G_i\}$ consisting of Pauli operators. We take the generalized $d$-level Pauli operators to be 
\begin{align}
D_{(x|z)} = \omega^{2^{-1}xz}X^xZ^z,  x,z \in \mathbb{Z}_d~,
\end{align}
where $2^{-1}$ is understood as the inverse of $2$ in $\mathbb{Z}_d$, and qudit $X$ and $Z$ are given by
\begin{align}
X\Ket{j} = \Ket{j+1}, \qquad Z\Ket{j} = \omega^j\Ket{j}
\end{align}
and $\omega = e^{2\pi i/d}$ is a $d$-th root of unity. 
The set of unitary operations that map Pauli operators to Pauli operators under conjugation is the Clifford group \cite{Clifford}, whose elements (for the single-qudit case) can be written as 
\begin{align}
C&= D_{(x|z)}U_F  \\
U_F &=\begin{cases} \frac{1}{\sqrt{d}}\sum_{j,k=0}^{d-1}{\omega^{2^{-1}\beta^{-1}(\alpha k^2-2jk+\delta j^2)}\Ket{j}\Bra{k}},\,  \beta \neq 0 \\
\sum_{k=0}^{d-1}\omega^{2^{-1}\alpha \gamma k^2}\Ket{\alpha k}\Bra{k}, \quad \beta = 0
\end{cases} \nonumber
\end{align}
where $F = \left(\begin{smallmatrix} \alpha & \beta \\ \gamma & \delta \end{smallmatrix}\right)$ is a symplectic matrix and all variables are elements of $\mathbb{Z}_d=\{0,1,\ldots,d-1\}$. 

Any magic state distillation protocol based on stabilizer codes consists of iteratively applying the following steps: 1. Prepare $n$ copies of the input state $\rho_{in}^{\otimes n}$; 2. Perform Pauli measurements corresponding to each of the $n-k$ generators $G_i$, and postselect on the desired outcome; 3. Optionally perform a Clifford transformation based on the measurement outcome. When successful, the output state(s) will be purified in the direction of the target magic state. 
We represent input states with depolarizing noise as 
\begin{align}
\rho_{M} = (1-p)\Ket{M}\Bra{M} +p\frac{\mathbb{1}_3}{d}~, \label{eqn:depol}
\end{align}
where $\Ket{M}$ is the target magic state, and $p^*$ will be used to denote the depolarizing noise rate above which our distillation routine no longer works.   

For qudits $(d>2)$ the subset of quantum states known to be classically simulable, and therefore not useful for magic state distillation, is prescribed by the Wigner polytope 
\begin{align*}
\text{Wigner polytope}:= \{ \rho : \Tr(\rho A_{x,z}) \geq 0,\quad  x,z \in \mathbb{Z}_d \}, 
\end{align*} 
where the phase point operators, for qutrits, are given by $A_{x,z} = D_{(x|z)}A_{0,0}D^\dag_{(x|z)}$ with
\begin{align}
A_{0,0} & = \left(\begin{array}{ccc} 1 & 0 & 0 \\ 0 & 0 & 1 \\ 0 & 1 & 0 \end{array}\right). \label{eqn:A00}
\end{align}
 This polytope contains the subset of quantum states that are nonnegatively represented in all $d^2$ positions of Gross' discrete Wigner function \cite{Gross:2006}
\begin{align}
W_\rho(x,z) = \frac{1}{d}\Tr(A_{x,z}\rho).
\end{align}
It turns out that the sum-negativity \cite{Victor2} of a state, 
\begin{align}
sn(\rho) = \sum_{x,z:W_\rho(x,z) < 0}|W_\rho(x,z)|,
\end{align}
is an operationally useful quantification of how far from the Wigner polytope a state is. 

\begin{figure}[h!]
\centering
\subfigure[]{
\includegraphics[scale=0.2]{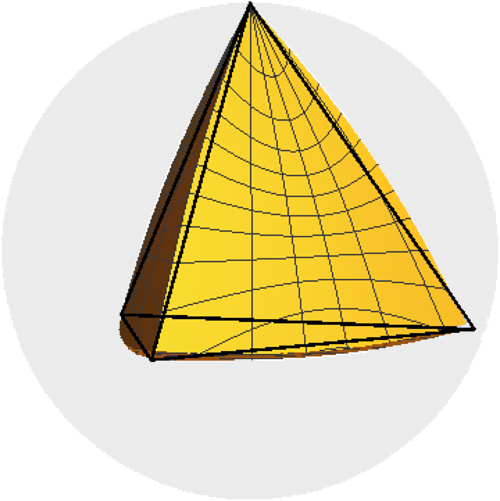}} \hspace{1cm}
\subfigure[]{
\includegraphics[scale=0.2]{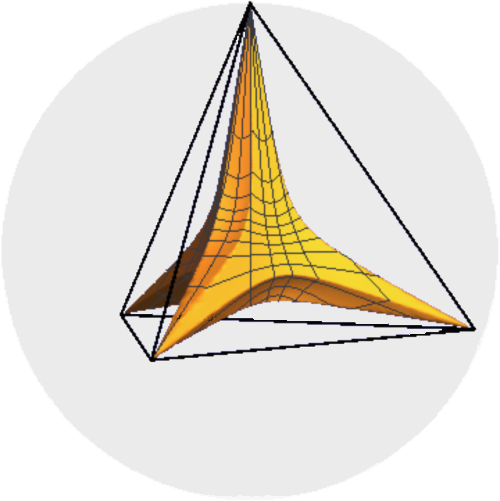}}
\caption{\label{fig:BS} Depolarized versions of qutrit states $ (\cos\theta,e^{i\phi}\sin\theta/\sqrt{2},e^{i\phi}\sin\theta/\sqrt{2})$: Points on surface of the sphere represent pure states, while every point in the interior is a depolarized version of the nearest surface state (see \eqref{eqn:depol}). Figure (a) depicts the set of states that are non-negatively represented in the Wigner function and hence useless for MSD. The volume roughly corresponds to a ``curvy tetrahedron'' which hereafter will be referred to as the Wigner tetrahedron. Figure (b) depicts the set of states expressible as convex combinations of stabilizer states. A regular tetrahedron is shown as a visual aid and the pure states at its vertices are stabilizer states $\{\Ket{0},(1,\omega^k,\omega^k)/\sqrt{3}, k\in \mathbb{Z}_3\}$.
}
\end{figure}

\textit{(a,b,b) subspace.$-$}We describe magic state distillation protocols that iteratively distill towards pure states within the $+1$-eigenspace of the $A_{0,0}$ phase point operator \eqref{eqn:A00}. This eigenspace is degenerate and has eigenvectors of the form $(a,b,b) \in \mathbb{C}^3$. Before describing the protocols, we give an overview of the geometrical interpretation of this eigenspace. We may parameterize pure states 
via
\begin{align}
\Ket{\psi} = (a,b,b) = (\cos\theta,e^{i\phi}\sin\theta/\sqrt{2},e^{i\phi}\sin\theta/\sqrt{2})
\label{eqn:PureState}
\end{align}
where $\phi \in [0,2\pi)$ and $\theta \in [0,\pi/2]$. The set of pure states corresponds to the surface of a sphere in analogy with the Bloch sphere for qubits. 
A point within the interior of the sphere with spherical coordinates $(r,\theta,\phi)$ corresponds to the state
\begin{align}
\rho = r\Ket{\psi(\theta,\phi)}\Bra{\psi(\theta,\phi)} + (1-r)\frac{\mathbb{1}_3}{d}
\end{align}
where $\Ket{\psi(\theta,\phi)}$ is given by \eqref{eqn:PureState}. 
However unlike the qubit Bloch sphere, states in the interior no longer correspond to convex combinations of surface states in general. For example, in the qubit Bloch sphere picture, we expect an equal mixture of any two diametrically opposite points to correspond to the maximally mixed state. However in our representation, mixing the North pole $\Ket{0}$ with the South pole $\Ket{N}=(\Ket{1}+\Ket{2})/\sqrt{2}$ gives
\begin{align}
\frac{1}{2}\Ket{0}\!\Bra{0}+\frac{1}{2}\Ket{N}\!\Bra{N}=\frac{1}{4}\left(\begin{array}{ccc} 2 & 0 &0 \\ 0& 1 & 1 \\0 & 1 & 1\end{array}\right) \neq \frac{\mathbb{1}_3}{3}.
\end{align}
Despite the fact that our representation does not respect convexity, we feel it provides good intuition for the relevant geometry and symmetries. Mixtures of stabilizer states as well as states with positive Wigner function form a closed volume within the sphere (see Fig.~\ref{fig:BS}). 

\begin{figure}
\includegraphics[scale=0.55]{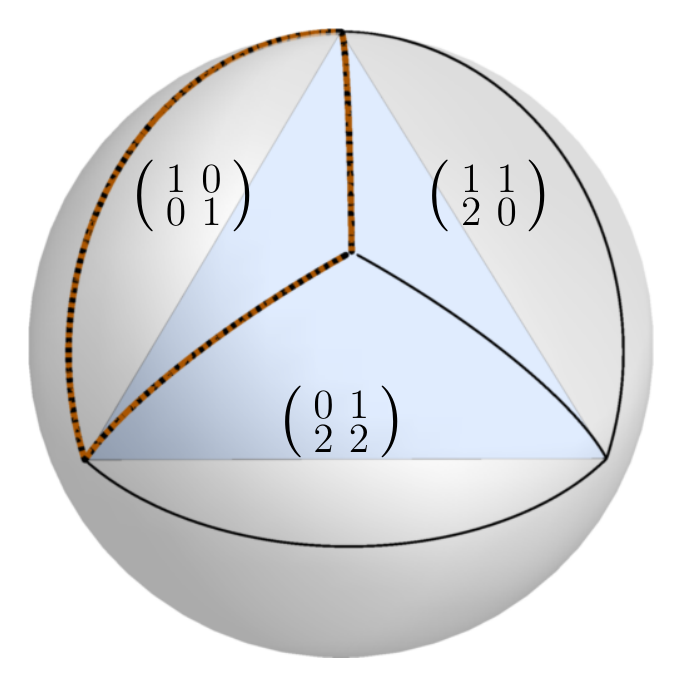}
\caption{\label{fig:Wedge} Clifford symmetries: The surface of the sphere lying outside one face of the Wigner tetrahedron is outlined in black and divided into 3 Clifford-equivalent wedges. Each wedge is labelled by the symplectic matrix $F$ such that $U_F$ maps states in that region back to the region of interest (highlighted).}
\end{figure}

In order to map the distillable region within the $(a,b,b)$ subspace, it will suffice to partition the space into Clifford-equivalent sections and find the distillable region for only one such section. It is a well-known property of the Clifford group that $A_{0,0}$ is invariant under the symplectic part of the Clifford group i.e., $A_{0,0}=U_F A_{0,0} U_F^\dag$ for all $F\in \textrm{SL}(2,\mathbb{Z}_d)$. 
Not only does the symplectic unitary $U_{\mathbb{1}_2}=\mathbb{1}_3$ fix every qutrit vector but the symplectic unitary $U_{-\mathbb{1}_2}=A_{0,0}$ also fixes every vector of the form $(a,b,b)$. Consequently, the set of non-trivial symplectic transformations acting on $(a,b,b)$ states is $\textrm{SL}(2,\mathbb{Z}_3)/\pm \mathbb{1}_2 = \textrm{PSL}(2,\mathbb{Z}_3)$. This latter symmetry group is isomorphic to the rotation group of the tetrahedron (i.e. the alternating group $A_4$).
Therefore we expect the entire space to partition into $|\textrm{PSL}(2,\mathbb{Z}_3)| = 12$ Clifford-equivalent regions. These regions correspond to the 4 Wigner tetrahedron faces, further divided into 3 wedges each as shown in Fig.~\ref{fig:Wedge}. 
We will consider distillation confined to the region highlighted in Fig.~\ref{fig:Wedge}. 
    
\textit{MSD routines.$-$} In this section we present two classes of magic states within the $(a,b,b)$ subspace, give explicit distillation schemes for each, and sketch out the corresponding distillable region. Magic states within the $(a,b,b)$ subspace can be generally split into two types, those lying outside the six Wigner tetrahedron edges, and those lying outside the four Wigner tetrahedron faces. By numerically searching over a large set of stabilizer codes, we find that there exists many limiting states outside the Wigner tetrahedron edges, where limiting state is taken to mean the end point of the iterative procedure based on a given stabilizer code.
\begin{figure}
\includegraphics[scale=.23]{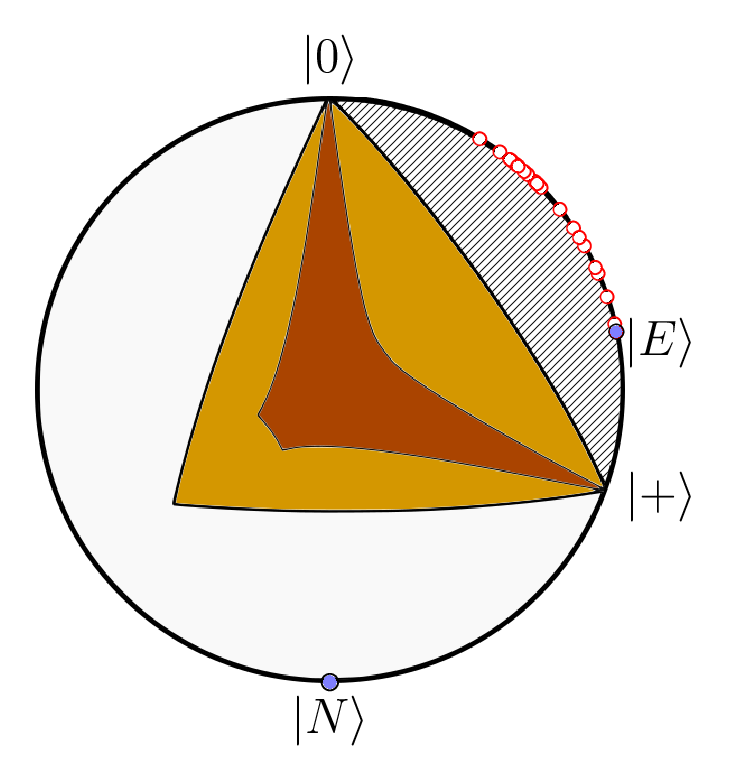}
\caption{\label{fig:Edge} Cross section through the $x$-$z$ plane showing limiting states of various MSD routines: Different circles correspond to different choices of stabilizer code and every such code can distill all states in the hatched area. Two magic states of interest are denoted by filled circles; the $\Ket{E}$ state lies on the arc joining $\ket{0}$ and $\ket{+}$, while the south pole $\ket{N}$ is Clifford equivalent to the Norrell state $\Ket{N^\prime}$ that we discuss in the text. The colored regions represent undistillable states inside the Wigner polytope (lighter, larger) and mixtures of stabilizer states (darker, smaller) i.e., this shows a 2-$d$ slice through Fig.~\ref{fig:BS}. }
\end{figure} 
In contrast, we find only one limiting state outside the Wigner tetrahedron faces, and this is the state geometrically located the furthest outside the Wigner tetrahedron in the middle of each face. These so-called Norrell states \cite{Victor2}, which were not previously known to be directly distillable, have maximal sum-negativity $sn=\frac{1}{3}$ arising from a Wigner function of $-\frac{1}{6}$ in two phase space points. 
 The set of numerically identified limiting states is shown in Fig.~\ref{fig:Edge}. 

\begin{figure}[h!]
\centering
\subfigure[\ The wedge of interest.]{
\includegraphics[trim=0cm 0.05cm 0cm 0cm,clip=true,scale=0.4]{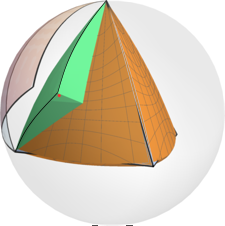}}
\subfigure[\ The $x$-$z$ plane.]{
\includegraphics[scale=0.18]{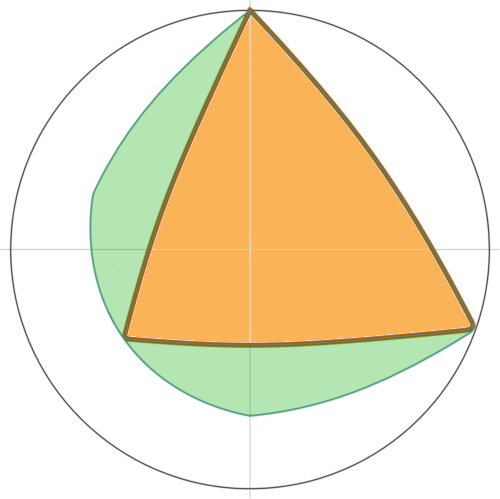}}\\
\subfigure[\ The $y$-$z$ plane.]{
\includegraphics[scale=0.18]{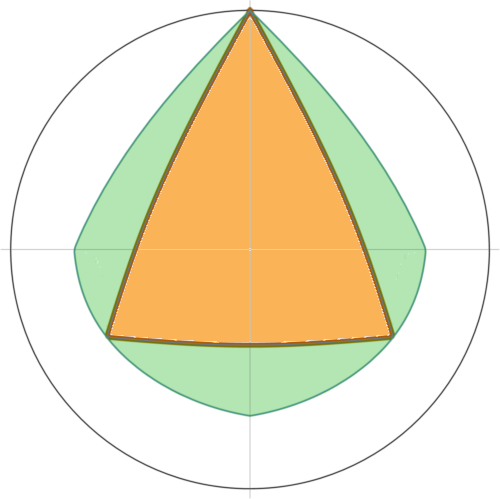}}
\subfigure[\ The $x$-$y$ plane.]{
\includegraphics[scale=0.18]{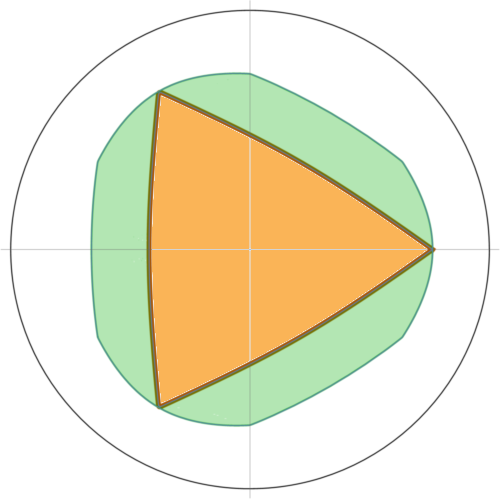}}
\caption{\label{fig:DistReg} The remaining undistillable region highlighted in green, after using the edge code \eqref{eqn:EdgeCode}. The small red dot in (a) corresponds to the threshold point for distilling $\Ket{N^\prime}$ using the face code \eqref{eqn:FaceCode}.}
\end{figure}

We now demonstrate distillation schemes based on stabilizer codes for both types of magic state. We begin with the edge states for which tight distillation thresholds were achieved, and then present the Norrell state which improves the distillation region outside the Wigner tetrahedron faces slightly. Out of the many edge-type magic states depicted in Fig.~\ref{fig:Edge}, the state whose code has the best overall distillable region will be presented. The state $\Ket{E} = (0.774149,0.447601,0.447601)$
which lies on the $x$-$z$ plane is distilled by a $[[4,1,2]]_3$ code given by generators:    
\begin{align}
\text{Edge,}\Ket{E}:\qquad \begin{tabular}{|c|cccc|cccc|} \hline
$G_1$ &0 & 0 & 0 & 2  &2 & 2 & 0 & 0 \\
$G_2$ &1 & 1 & 0 &1 &1 & 1 & 2 & 2 \\
$G_3$ &0 & 0 & 1 & 0  &2 & 0 & 0 & 0 \\
$Z_L$ &2 & 0 & 0 & 2  &2 & 2 & 1 & 2 \\
$X_L$ &0 & 0 & 0 & 0  &1 & 2 & 0 & 0 \\ \hline
\end{tabular} \label{eqn:EdgeCode}
\end{align}   
in $(x|z)$ notation where each element is taken to be a generalized Pauli operator. This magic state $\Ket{E}$ may be transformed into an equatorial state useful for state injection by first applying a symplectic unitary $U_F$ with $F=\left(\begin{smallmatrix} 1 & 1\\0 & 1 \end{smallmatrix}\right)$
and then following the parity check and equatorialization procedures as outlined in \cite{CAB1}.
For all states $\ket{\theta}:=(\cos \theta, \sin \theta/\sqrt{2},\sin \theta/\sqrt{2})$ along the Wigner tetrahedron edge, on the line $(x,y,z) = (\sin 2\theta,0,\cos 2\theta)$ with $0\leq \theta \leq \arccos 1/\sqrt{3}$, we find an error threshold given by
\begin{align}
p^* = 1-\frac{4}{1+3\cos 2\theta+3\sqrt{2}\sin 2\theta}~,
\end{align}
which corresponds to a state $\rho_\theta = (1-p^*)\Ket{\theta}\Bra{\theta}+p^*\mathbb{1}_3/3$ with Wigner function
\begin{align}
W_{\rho_\theta} &=\left(\begin{array}{ccc}
r & s & s \\ 
t & 0 &0\\
t & 0 & 0\\
\end{array}\right)\quad\begin{array}{c}
r=\frac{\cos 2\theta+\sqrt{2}\sin 2\theta+3}{9\cos 2\theta+9\sqrt{2}\sin 2\theta+3} \geq 0\\
s=\frac{4\cos 2\theta+\sqrt{2}\sin 2\theta}{9\cos 2\theta+9\sqrt{2}\sin 2\theta+3} \geq 0\\
t=\frac{\sqrt{2}}{3\cot 2\theta+\csc 2\theta+3\sqrt{2}}  \geq 0 \end{array}
\end{align}
and therefore distillation is tight to the Wigner polytope boundary for all points along the Wigner tetrahedron edges. The maximally robust edge state occurs at $\theta = \frac{1}{2}\cos^{-1}(\frac{1}{\sqrt{3}})$ and this corresponds to the $+1$ eigenstate of the qutrit Fourier transform. 
Using our edge code, this state can tolerate depolarizing noise up to 
\begin{align}
p^* = 1 - \frac{4}{1+3\sqrt{3}} \approx 0.354438~,
\end{align} 
which is the best known depolarizing noise threshold for qutrits. The edge code distills points along the depolarizing axis of the Norrell state  $\Ket{N^\prime} = (2,-1,-1)/\sqrt{6}$ with a threshold of $p^* \approx 0.304379$.
The entire distillable region of this $[[4,1,2]]_3$ edge code confined to the Clifford-equivalent wedge of interest is shown in Fig.~\ref{fig:DistReg}. The situation is analogous to the qubit picture wherein distillation is tight for all edges ($H$-type) and there is a pocket of undistillable states outside the Wigner tetrahedron faces ($T$-type).

%

The distillation region may be improved slightly by a second stabilizer code which has the Norrell state  $\Ket{N^\prime} = (2,-1,-1)/\sqrt{6}$ as a limiting state. The Norrell state $\ket{N^\prime}$ is Clifford equivalent to the south pole state $\Ket{N}$ shown in Fig.~\ref{fig:Wedge}. This $[[4,1,2]]_3$ code has generators:
\begin{align}
\text{Face,}\Ket{N^\prime}:\qquad \begin{tabular}{|c|cccc|cccc|} \hline
$G_1$ &2 & 0 & 0 & 2  &1 & 2 & 0 &1 \\
$G_2$ &2 & 1 & 0 & 1 &1 & 0 & 1 & 0 \\
$G_3$ &1 & 0 & 1 & 2  &0 & 2 & 1 & 0 \\
$Z_L$ &1 & 0 & 0 & 2  &1 & 0 & 1 & 2 \\
$X_L$ &0 & 0 & 0 & 0 &1 & 2 & 1 & 2 \\ \hline
\end{tabular} \label{eqn:FaceCode} 
\end{align}   
We find that the Norrell state $\Ket{N^\prime}$ is distilled up to a threshold noise rate of $p^* \approx 0.32989$, 
which increases the distillable region as shown in Fig.~\ref{fig:DistReg}. The equatorialization procedure of \cite{CAB1} converts the Norrell state into something useful for state injection (two copies of $\Ket{N^\prime}$ enable implementation of the non-Clifford unitary $U=\mathrm{diag}(1,1,-1)$  \cite{CAB1}).

The success probabilities $p_\text{succ}=\Tr(\rho_{in}^{\otimes 4} \Pi_\textsc{C})$ for both codes for points along the depolarizing axes of their respective limiting states begin at approximately $p_\text{succ} = 0.12$ for $p = 0$, 
assuming post-selection on the trivial error syndrome. 
Similar to \cite{CAB1} error suppression appears to be linear, which means that despite the small code size, both codes are outperformed in terms of efficiency by a previously known qutrit code \cite{CAB2,CAB3}. Our focus was on maximizing the range of applicability of MSD schemes rather than their efficiency. Nevertheless, exploring the landscape of stabilizer codes with non-stabilizer limiting states may have practical consequences if these codes can be adapted or used in conjunction with others (as in \cite{Meier:2013}) to produce more efficient schemes. 
%
 
 \textit{Summary \& open questions.$-$} We have demonstrated the tightness of magic state distillation schemes to the Wigner polytope boundary for $d=3$ in some directions. This represents a first step towards showing that all negatively represented or contextual states are distillable by some routine. Our results help to map out the set of codes that are useful for MSD, over and above those known to useful due to their transversality properties \cite{BK,CAB2,Bravyi:2012,Jones:2013}, and the set of limiting states for higher dimensional systems. We have shown that a certain class of maximally non-stabilizer state is distillable by at least one stabilizer code. We have given a convenient parameterization of the +1-eigenspace of the $A_{0,0}$ phase space operator which allows us to easily visualize the distillable region and relevant symmetries. The distillable region in our work presents an interesting analogy to known results for the qubit Bloch sphere, where all states outside the edge of the relevant polytope are distillable. Why these states in particular are limiting states of an MSD routine, and what properties of the identified stabilizer codes make them useful for magic state distillation are interesting open questions.

\textit{Acknowledgments.$-$} We thank Earl Campbell for helpful comments. The authors acknowledge financial support from the Government of Canada through NSERC through the CGS-M program and the discovery grant program, as well as the U.~S.~ Army Research Office through grant W911NF-14-1-0103, and FQXI.


\begin{thebibliography}{99}
\bibitem{Shor}
P.~Shor, ``Fault-tolerant quantum computation'' in \textit{Foundations of Computer Science, 1996.~Proceedings., 37th Annual Symposium on} (IEEE, 1996), pp.~56-65.

\bibitem{Kit}
A.~Kitaev, 
\newblock ``Fault-tolerant quantum computation by anyons''
\href{http://dx.doi.org/10.1016/S0003-4916(02)00018-0}
{Ann.~Phys.~\textbf{303}, 2 (2013).}

\bibitem{Raus}
R.~Raussendorf and J.~Harrington, 
\newblock ``Fault-Tolerant Quantum Computation with High Threshold in Two Dimensions''
\href{http://dx.doi.org/10.1103/PhysRevLett.98.190504}
{Phys.~Rev.~Lett.~\textbf{98}, 190504 (2007).}


\bibitem{Gottesman1}
D.~Gottesman, 
\newblock ``Theory of fault-tolerant quantum computation''
\href{http://dx.doi.org/10.1103/PhysRevA.57.127}
{Phys.~Rev.~A \textbf{57}, 127 (1998).}


\bibitem{Paetznick:2013}
A.~Paetznick and B.~W.~Reichardt,
\newblock ``Universal Fault-Tolerant Quantum Computation with Only Transversal Gates and Error Correction''
\href{http://dx.doi.org/10.1103/PhysRevLett.111.090505}
{Phys.~Rev.~Lett.~\textbf{111}, 090505 (2013).}

\bibitem{TJOC:2013}
T.~Jochym-O’Connor and R.~Laflamme,
\newblock ``Using Concatenated Quantum Codes for Universal Fault-Tolerant Quantum Gates''
\href{http://dx.doi.org/10.1103/PhysRevLett.112.010505}
{Phys.~Rev.~Lett.~\textbf{112}, 010505 (2014).}


\bibitem{Knill:2005}
E.~Knill,
\newblock ``Quantum computing with realistically noisy devices''
\href{http://dx.doi.org/10.1038/nature03350}
{Nature \textbf{434}, 39 (2005).}

\bibitem{BK}
S.~Bravyi and A.~Kitaev,
\newblock ``Universal quantum computation with ideal Clifford gates and noisy ancillas''
\href{http://dx.doi.org/10.1103/PhysRevA.71.022316} 
{Phys.~Rev.~A \textbf{71}, 022316 (2005).}

\bibitem{Reichardt}
B.~W.~Reichardt,
\newblock ``Quantum Universality from Magic States Distillation Applied to CSS Codes''
\href{http://dx.doi.org/10.1007/s11128-005-7654-8} 
{Quantum Information Processing \textbf{4}, 251 (2005).}

\bibitem{CB1}
E.~T.~Campbell and D.~E.~Browne,
\newblock ``Bound States for Magic State Distillation in Fault-Tolerant Quantum Computation''
\href{http://dx.doi.org/10.1103/PhysRevLett.104.030503}
{Phys.~Rev.~Lett.~\textbf{104}, 030503 (2010).}

\bibitem{CB2}
E.~T.~Campbell and D.~E.~Browne, 
\newblock ``On the Structure of Protocols for Magic State Distillation''
\href{http://dx.doi.org/10.1007/978-3-642-10698-9_3}
{Lecture Notes in Computer Science \textbf{5906}, 20 (2009).}

\bibitem{Wootters:1987}
W.~K.~Wootters,
\newblock ``A Wigner-function formulation of finite-state quantum mechanics''
\newblock \href{doi:10.1016/0003-4916(87)90176-X}{Ann.~Phys.~\textbf{176}, 1 (1987).}


\bibitem{Gross:2006}
D.~Gross,
``Hudson’s theorem for finite-dimensional quantum systems''
\newblock \href{http://dx.doi.org/10.1063/1.2393152}{J.~Math.~Phys.~\textbf{12} 47 122107 (2006).}

\bibitem{Victor}
V.~Veitch, C.~Ferrie, D.~Gross, and J.~Emerson, 
\newblock ``Negative quasi-probability as a resource for quantum computation ''
\href{http://dx.doi.org/10.1088/1367-2630/14/11/113011}
{New J.~Phys.~\textbf{14}, 113011 (2012).}

\bibitem{Victor2}
V.~Veitch, S.~A.~H.~Mousavian, D.~Gottesman, and J.~Emerson, 
\newblock ``The resource theory of stabilizer quantum computation ''
\href{http://dx.doi.org/10.1088/1367-2630/16/1/013009}
{New J.~Phys.~\textbf{16}, 013009 (2014).}


\bibitem{Context}
M.~Howard, J.~Wallman, V.~Veitch, and J.~Emerson,
\newblock ``Contextuality supplies the ‘magic’ for quantum computation''
\href{http://dx.doi.org/10.1038/nature13460}
{Nature \textbf{510}, 351–355 (2014).}


\bibitem{CAB1}
H.~Anwar, E.~T.~Campbell, and D.~E.~Browne,
\newblock ``Qutrit magic state distillation ''
\href{http://dx.doi.org/10.1088/1367-2630/14/6/063006}
{New J.~Phys.~\textbf{14}, 063006 (2012).}

\bibitem{CAB2}
E.~T.~Campbell,
\newblock ``Enhanced Fault-Tolerant Quantum Computing in $d$-Level Systems''
\href{http://dx.doi.org/10.1103/PhysRevLett.113.230501}{Phys.~Rev.~Lett.~
\textbf{113}, 230501, (2014).}

\bibitem{CAB3}
E.~T.~Campbell, H.~Anwar, and D.~E.~Browne,
\newblock ``Magic-State Distillation in All Prime Dimensions Using Quantum Reed-Muller Codes''
\href{http://dx.doi.org/10.1103/PhysRevX.2.041021}
{Phys.~Rev.~X \textbf{2}, 041021 (2012).}





\bibitem{Clifford}
D.~M.~Appleby, I.~Bengtsson, and S.~Chaturvedi,
\newblock ``Spectra of phase point operators in odd prime dimensions and the extended Clifford group ''
\href{http://dx.doi.org/10.1063/1.2824479}
{J.~Math.~Phys.~\textbf{49}, 012102 (2008).}
%

\bibitem{Meier:2013}
A.~M.~Meier, B.~Eastin, E.~Knill, 
\newblock ``Magic-state distillation with the four-qubit code''
\href{http://arxiv.org/abs/1204.4221}
{Quantum Information \& Communication \textbf{13}, 195 (2013).} 

\bibitem{Bravyi:2012}
S.~Bravyi, and J.~Haah,
\newblock ``Magic-state distillation with low overhead''
\href{http://dx.doi.org/10.1103/PhysRevA.86.052329}
{Phys.~Rev.~A \textbf{86}, 052329 (2012).} 

\bibitem{Jones:2013}
C.~Jones,
\newblock ``Multilevel distillation of magic states for quantum computing''
\href{http://dx.doi.org/10.1103/PhysRevA.87.042305}
{Phys.~Rev.~A \textbf{87}, 042305 (2013).} 


\end{thebibliography}
\end{document}